\documentclass[twoside]{dis09}
\usepackage[latin1]{inputenc}
\usepackage[dvips]{graphicx,epsfig,color}
\usepackage{wrapfig,rotating}
\usepackage{amssymb,amsmath,array}

\begin{document}
\title{Search for New Physics in the $\mathbf{B_s}$ Sector at the Tevatron}

\author{Thomas Kuhr\\ on behalf of the CDF and D0 Collaborations
%
%
\vspace{.3cm}\\
%
Institut f\"ur Experimentelle Kernphysik - KIT \\
Wolfgang-Gaede-Str.\ 1, 76131 Karlsruhe - Germany
%
}

\maketitle

\begin{abstract}
While $B^0$ and $B^+$ mesons are well studied at B-factories, large samples
of $B^0_s$ mesons are only available at the Tevatron so far.
Since the $B^0_s$ meson consists of quarks of the second and third generation
it provides a complementary probe for searches for new physics effects.
Results of the CDF and D0 experiments on the decays $B^0_s \rightarrow \mu^+\mu^-$
and $B^0_s \rightarrow \mu^+\mu^-\phi$ as well as on \textit{CP} violation
in $B^0_s$ mixing and in interference between mixing and decay are presented.
\end{abstract}

\section{Introduction}
At the Tevatron $B^0_s$ mesons are copiously produced in $p\bar{p}$ collisions
at a center of mass energy of $\sqrt{s}=1.96$ TeV.
Having recorded datasets corresponding to an integrated luminosity of more
than 5 fb$^{-1}$, the two Tevatron experiments CDF and D0 are able to perform
detailed studies of $B^0_s$ mesons.
Essential for these analyses are triggers that efficiently select events with
$B^0_s$ mesons.
They are identified by decays to $J/\psi$ mesons with $J/\psi \rightarrow \mu^+\mu^-$,
semileptonic decays to muons (D0), or hadronic decays with a displaced decay
vertex (CDF).

The $B^0_s$ mesons are a promising place to search for physics beyond the
Standard Model (SM) because several flavor changing neutral current (FCNC) processes
can be measured in the $B^0_s$ system.
As there are no FCNC processes at tree level in the SM, new physics
contributions may be sizable.
Two rare FCNC decay modes of the $B^0_s$ that are analyzed at the Tevatron are
$B^0_s \rightarrow \mu^+\mu^-$ and $B^0_s \rightarrow \mu^+\mu^- \phi$.

Another interesting FCNC process is the oscillation of $B^0_s$ mesons.
It is described by a Schr\"odinger equation
\begin{equation*}
i\frac{d}{dt} \left( \begin{array}{c} \left|B_s^0(t)\right> \\ \left|\bar{B}_s^0(t)\right> \end{array} \right)
= \mathbf{M} - \frac{i}{2} \mathbf{\Gamma}
\left( \begin{array}{c} \left|B_s^0(t)\right> \\ \left|\bar{B}_s^0(t)\right> \end{array} \right)
\end{equation*}
where $\mathbf{M}$ and $\mathbf{\Gamma}$ are $2\times 2$ hermitian matrices.
The solution are two eigenstates with defined masses, $m_H$ and $m_L$, and decay widths,
$\Gamma_H$ and $\Gamma_L$, respectively.
While the measured oscillation frequency, $\Delta m=m_H-m_L$, agrees well with
the (less precise) SM prediction, there may still be sizable new physics
contributions to the phase $\phi=\arg(-M_{12}/\Gamma_{12})$.
This would result in a violation of the \textit{CP} symmetry in the $B^0_s$ system, that is 
expected to be conserved to very good approximation in the SM.

\section{$\mathbf{B^0_s \rightarrow \mu^+\mu^-}$}
In the SM the FCNC process $B^0_s \rightarrow \mu^+\mu^-$ is predicted to have a branching ratio 
of $\mathcal{B}(B^0_s \rightarrow \mu^+\mu^-) = (3.35 \pm 0.32) \times 10^{-9}$\ \cite{Blanke:2006ig}.
A significant enhancement of the branching ratio is expected in
several new physics models.
For example in the minimal super-symmetric standard model the $B^0_s$ branching ratio 
is proportional to $\tan^6\beta$ where $\tan\beta$ is the ratio between the vacuum expectation values
of the two neutral Higgs fields.
In $R$-parity violating super-symmetric models an enhancement is possible even at low 
values of $\tan\beta$.

\begin{figure}
\centering
\begin{minipage}{0.64\textwidth}
\includegraphics[width=\textwidth]{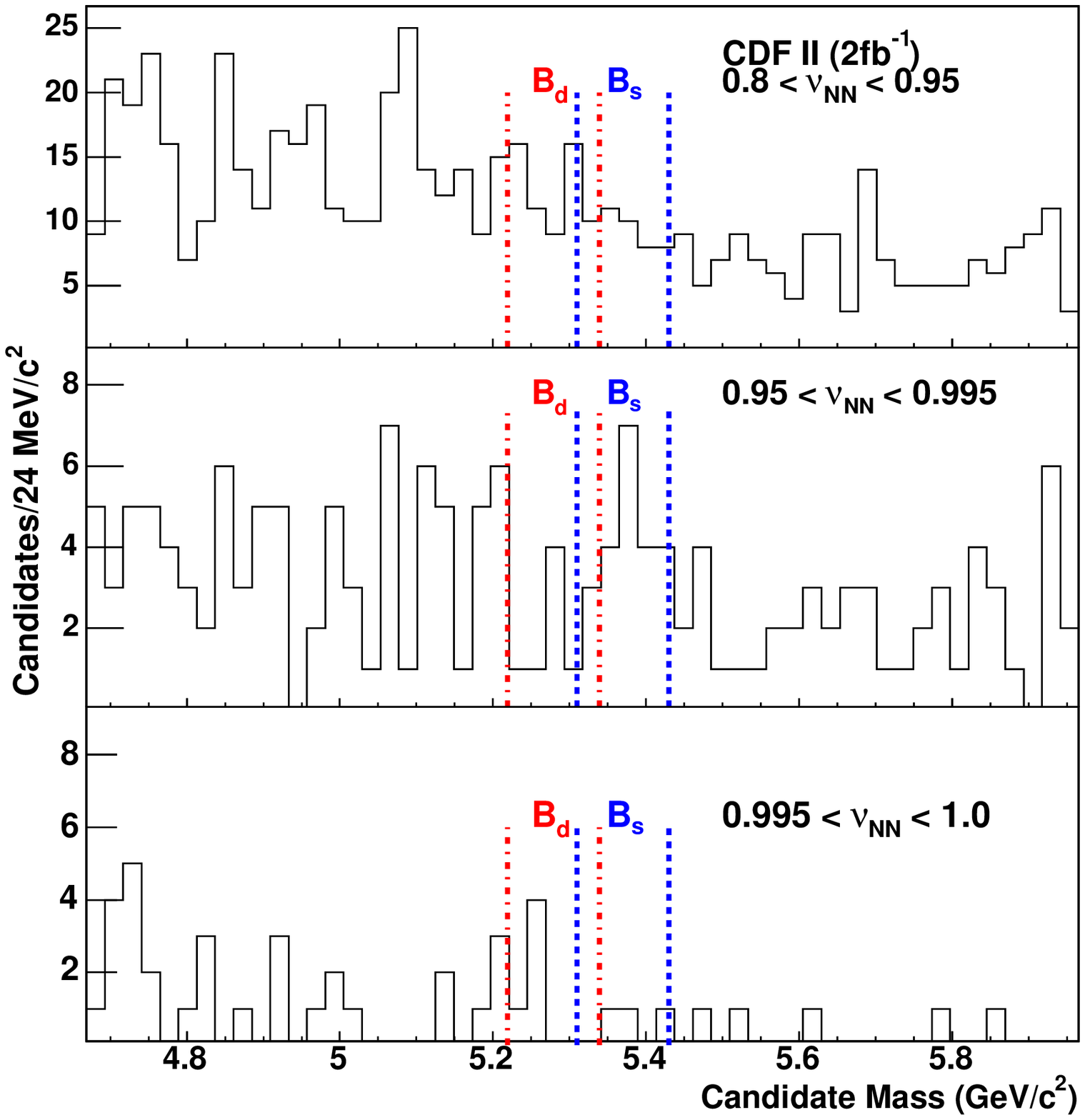}
\end{minipage}
\begin{minipage}{0.32\textwidth}
\includegraphics[width=\textwidth]{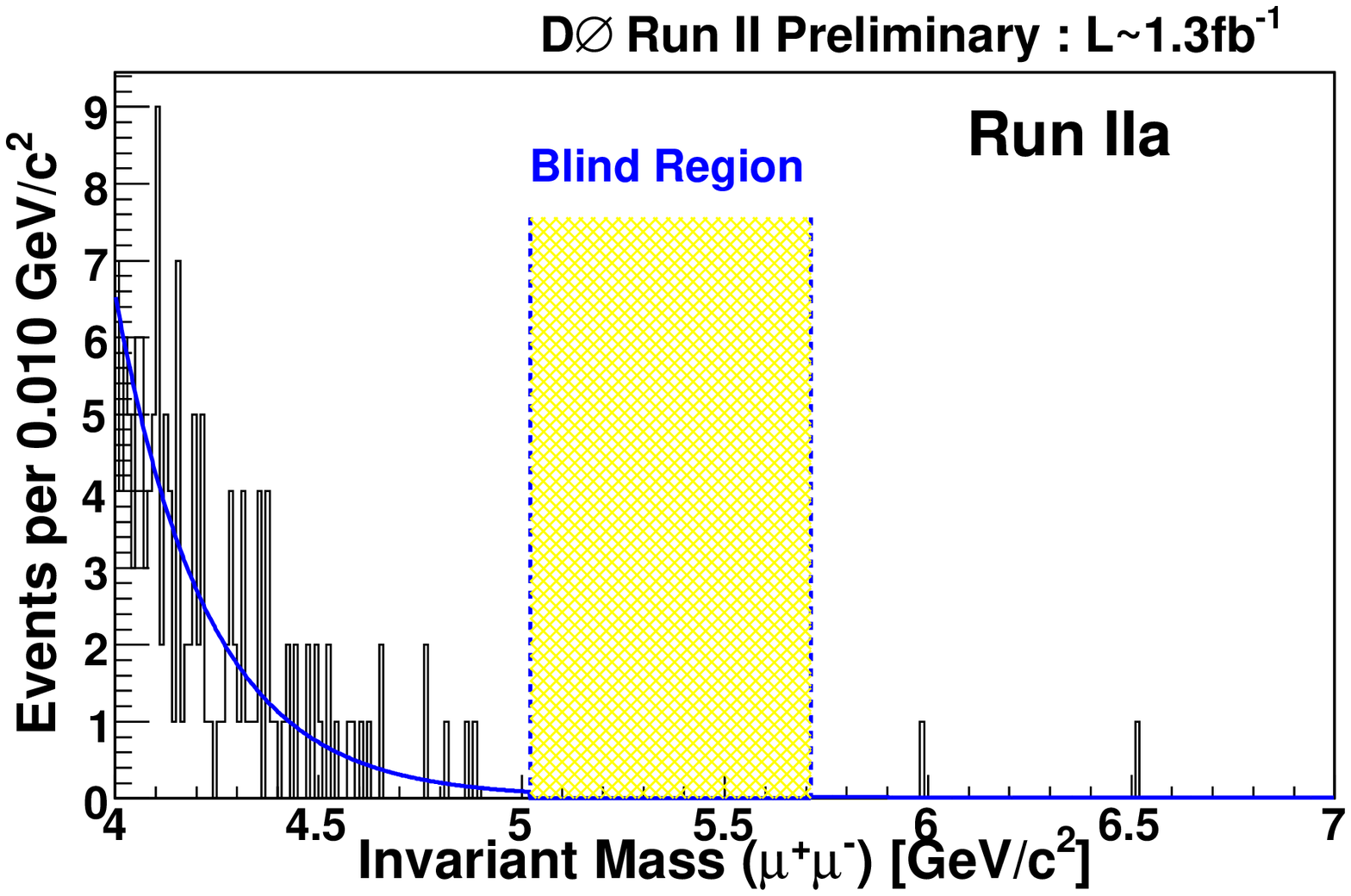}
\includegraphics[width=\textwidth]{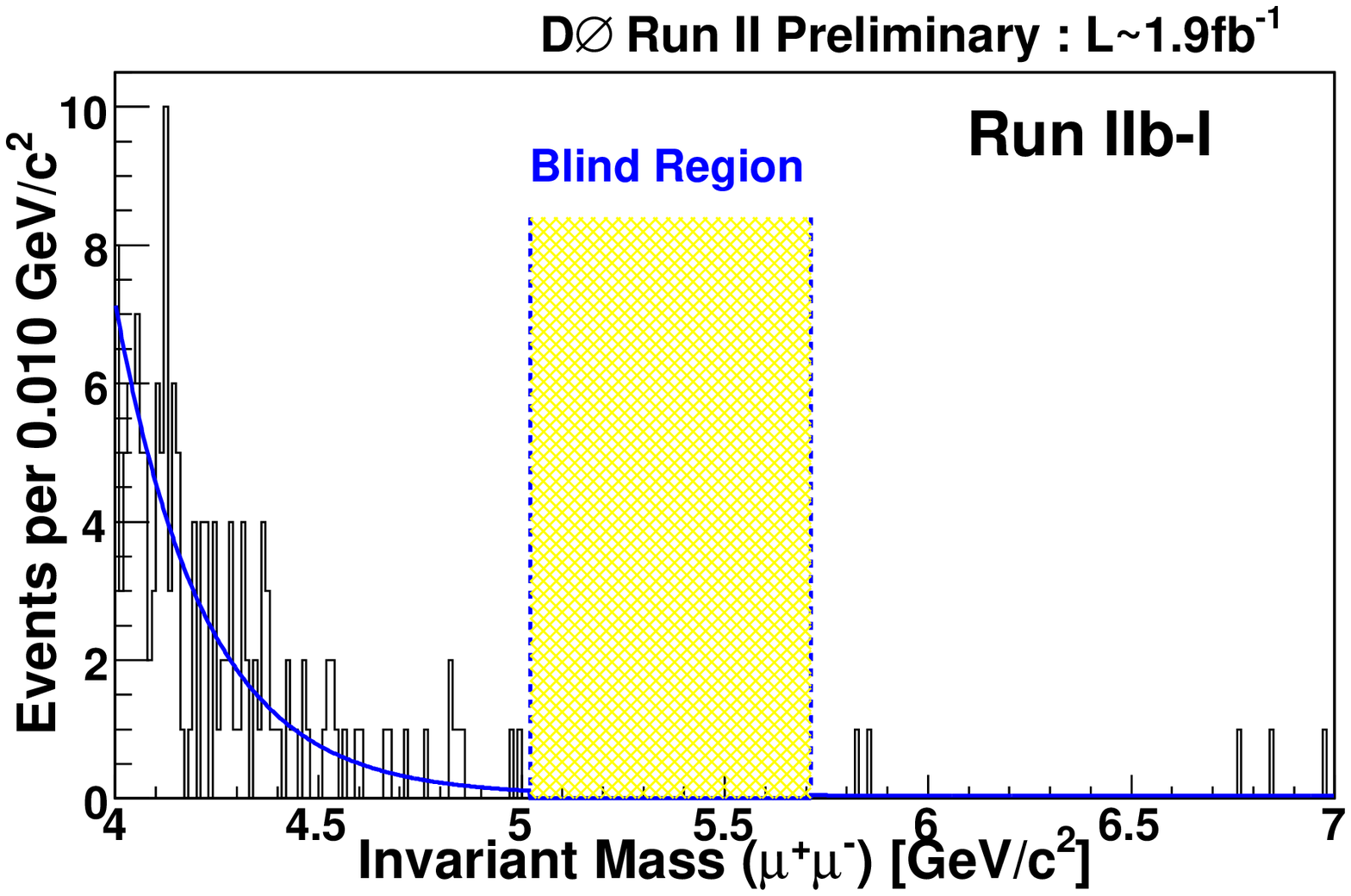}
\includegraphics[width=\textwidth]{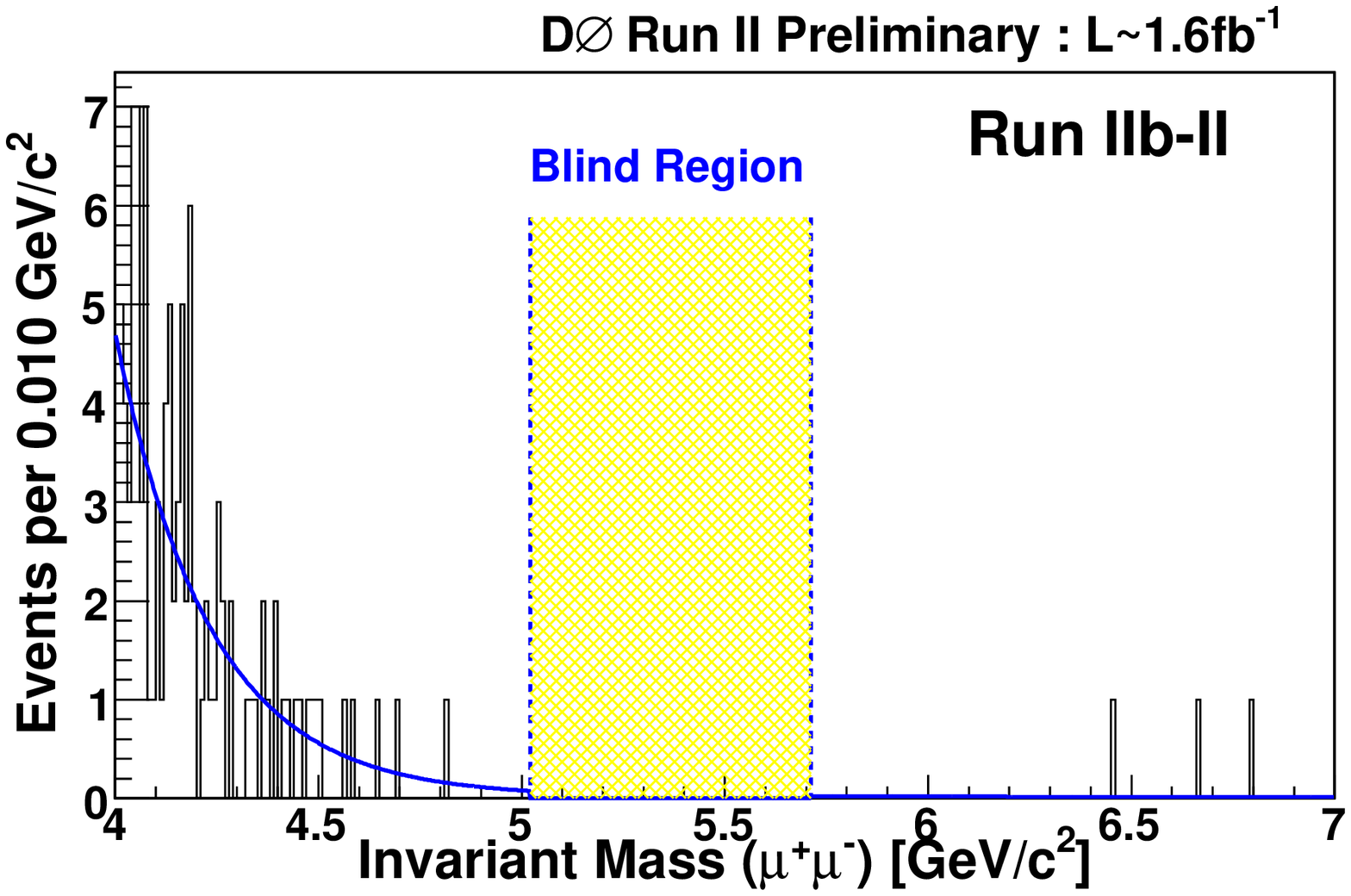}
\end{minipage}
\caption{Invariant mass spectrum of $B^0_s \rightarrow \mu^+\mu^-$ candidates
measured by CDF in three bins of neural network output (left) and 
by D0 in three run ranges (right).}
\label{fig:BsMuMu}
\end{figure}

Both Tevatron experiments optimize the selection of $B^0_s \rightarrow \mu^+\mu^-$ candidates 
using simulated signal events and background events from mass sidebands.
While D0 combines the discriminant variables in a boosted decision tree, CDF uses a neural network (NN).
The combinatorial background is estimated by a fit to the mass sidebands.
The contribution from decays of $B$ mesons to two light hadrons, which could peak in the signal
mass region, was estimated to be an order of magnitude lower than the combinatorial background.
To obtain an absolute branching ratio the number of signal events is normalized to the
high-statistics $B^+ \rightarrow J/\psi K^+$ mode.
For the limit calculation CDF splits the data sample in three bins in NN output and five bins in mass
which improves the sensitivity by 15\% compared to using just one bin.
The D0 analysis is performed on three separate data samples corresponding to three different
detector configurations.
CDF does not see a significant excess (Fig.\ \ref{fig:BsMuMu} left) in a data sample of 2 fb$^{-1}$
and sets a 90\% confidence level (CL) limit at 
$\mathcal{B}(B^0_s \rightarrow \mu^+\mu^-) < 4.7 \times 10^{-8}$\ \cite{Aaltonen:2007kv}.
D0 has not yet unblinded the signal region (Fig.\ \ref{fig:BsMuMu} right) and quotes an expected
upper limit at 90\% CL of $4.3 \times 10^{-8}$ for a data sample of 5 fb$^{-1}$\ \cite{D0:Bsmumu}.

\section{$\mathbf{B^0_s \rightarrow \mu^+\mu^- \phi}$}
Current measurements of the FCNC quark level process $b \rightarrow s \gamma$ in
$B^0$ and $B^+$ decays already provide stringent constraints on new physics models.
At the Tevatron this process can be analyzed in $B^0_s$ decays if the photon is virtual
and decays to $\mu^+\mu^-$.
At hadron level the decay $B^0_s \rightarrow \mu^+\mu^- \phi$ is reconstructed.
In the SM it is predicted to have a branching ratio of the order 
$\mathcal{B}(B^0_s \rightarrow \mu^+\mu^- \phi) = 1.6 \times 10^{-6}$\ \cite{Geng:2003su}.

\begin{figure}
\centering
\begin{minipage}{0.32\textwidth}
\includegraphics[width=\textwidth]{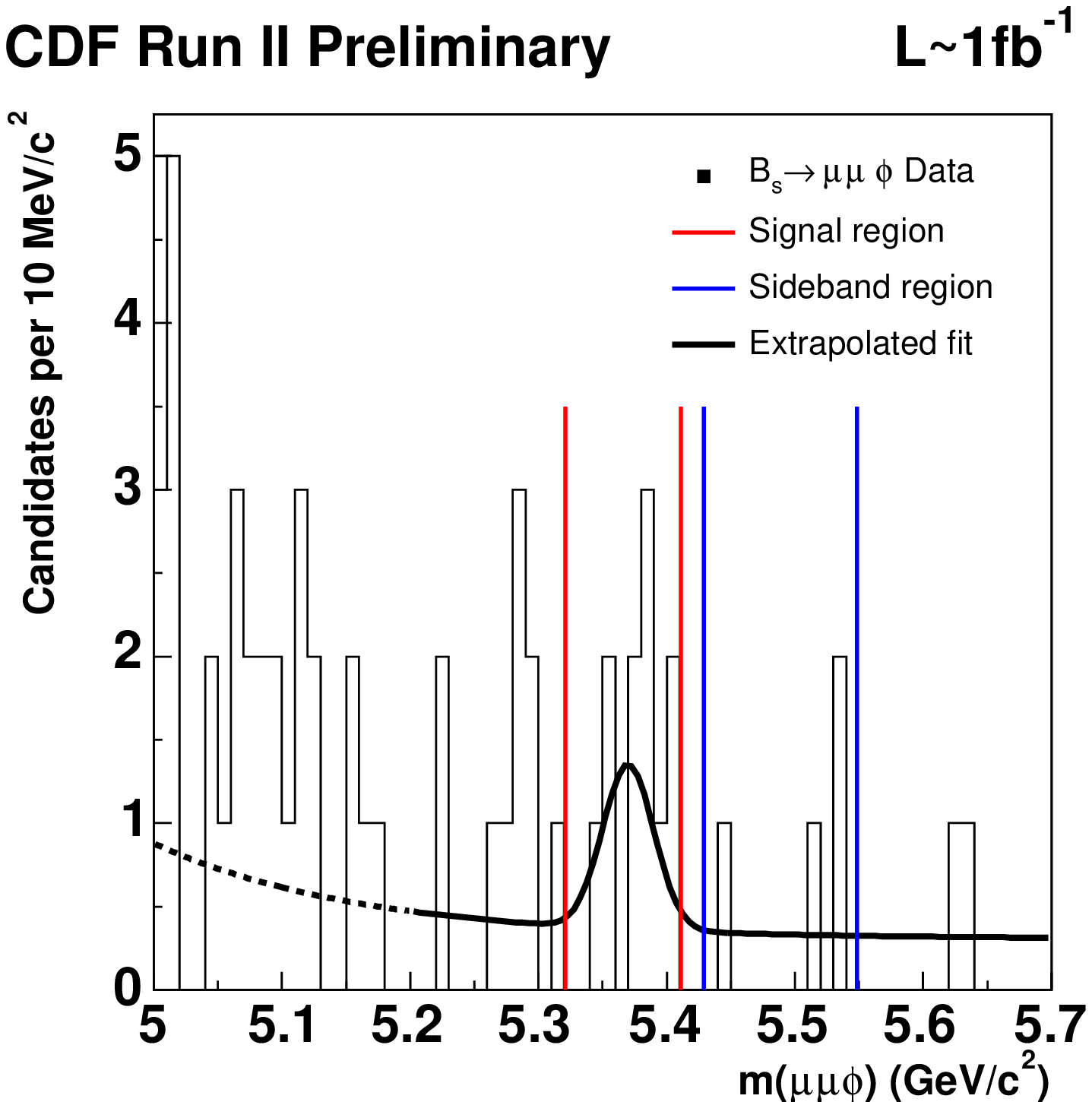}
\end{minipage}
\begin{minipage}{0.6\textwidth}
\includegraphics[width=\textwidth]{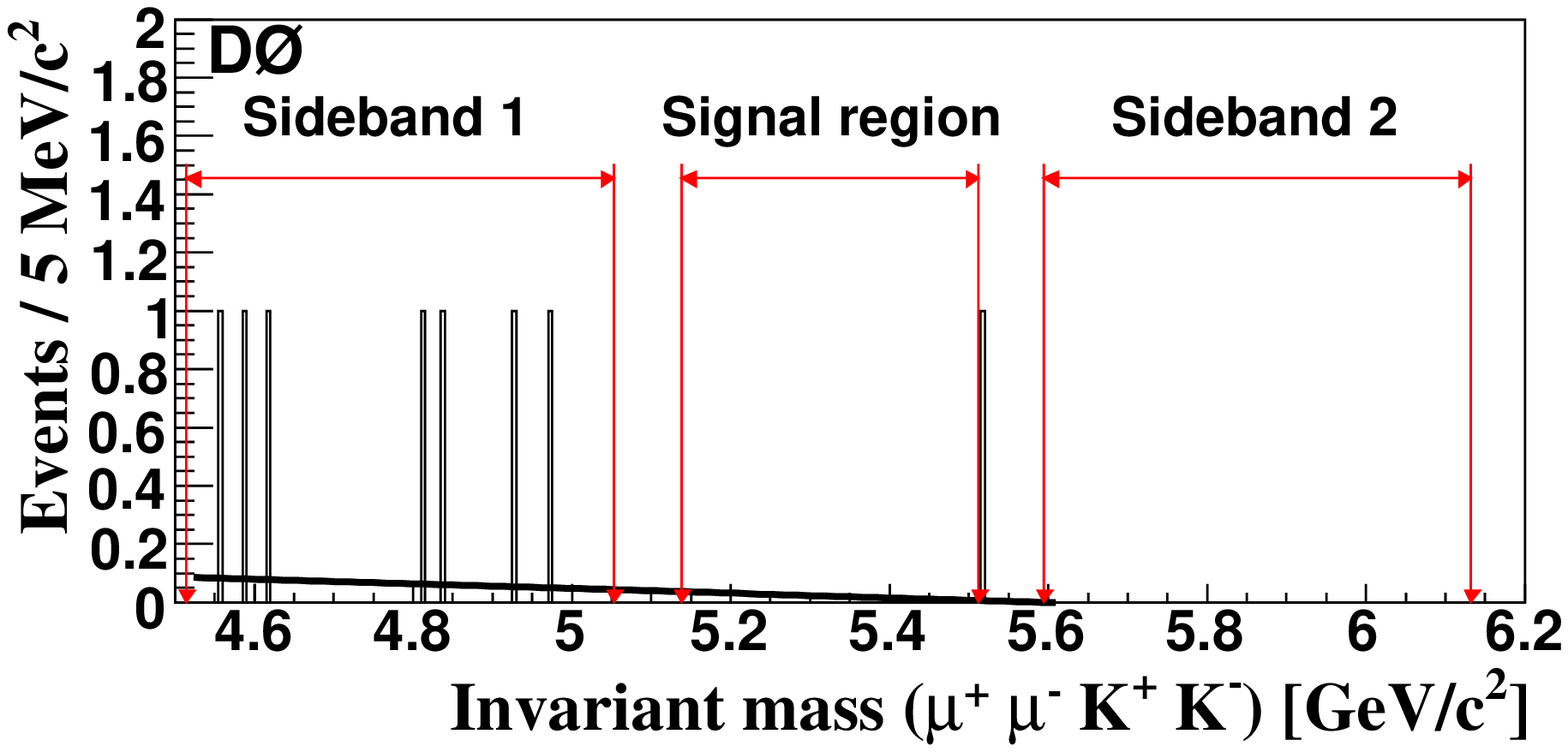}
\end{minipage}
\caption{Invariant mass spectrum of $B^0_s \rightarrow \mu^+\mu^-\phi$ candidates
measured by CDF (left) and D0 (right).}
\label{fig:Bsmumuphi}
\end{figure}

Decays to $J/\psi\,\phi$ and $\psi(2S)\,\phi$ are excluded from the measurement.
However, the former one is used as a normalization channel.
In a data sample of 0.92 fb$^{-1}$ CDF sees an excess with a significance of
2.4 standard deviations (Fig.\ \ref{fig:Bsmumuphi} left).
Assuming this excess comes from $B^0_s \rightarrow \mu^+\mu^- \phi$ decays,
a branching ratio of $\mathcal{B}(B^0_s \rightarrow \mu^+\mu^-\phi) = 
(1.70 \pm 0.82\mbox{(stat.)} \pm 0.64\mbox{(syst.)}) \times 10^{-6}$ is measured,
well consistent with the SM expectation.
Since the signal is not significant CDF quotes a 95\% CL limit on the relative
branching ratio of
$\mathcal{B}(B^0_s \rightarrow \mu^+\mu^-\phi)/\mathcal{B}(B^0_s \rightarrow J/\psi\,\phi)
< 2.6 \times 10^{-3}$, corresponding to a limit on the absolute branching ratio of
$\mathcal{B}(B^0_s \rightarrow \mu^+\mu^-\phi) < 6.0 \times 10^{-6}$\ \cite{Aaltonen:2008xf}.
D0 observes no events in the signal region in a data sample of 0.45 fb$^{-1}$ 
(Fig.\ \ref{fig:Bsmumuphi} right).
The derived 95\% CL upper limits are 
$\mathcal{B}(B^0_s \rightarrow \mu^+\mu^-\phi)/\mathcal{B}(B^0_s \rightarrow J/\psi\,\phi)
< 4.4 \times 10^{-3}$ and 
$\mathcal{B}(B^0_s \rightarrow \mu^+\mu^-\phi) < 4.1 \times 10^{-6}$\ \cite{Abazov:2006qm}.

\section{\textit{CP} violation in $\mathbf{B^0_s}$ mixing}
\textit{CP} violation in mixing can be observed by measuring decay rates to
flavor specific final states, $f$ and $\bar{f}$, that do not exhibit direct 
\textit{CP} violation.
The asymmetry
$a^s_{SL} = [\Gamma(\bar{B}^0_s \rightarrow f)-\Gamma(B^0_s \rightarrow \bar{f})]/
[\Gamma(\bar{B}^0_s \rightarrow f)+\Gamma(B^0_s \rightarrow \bar{f}]
= (\Delta\Gamma/\Delta m) \tan\phi$ with $\Delta\Gamma=\Gamma_L-\Gamma_H$
is expected to be $(2.06 \pm 0.57) \times 10^{-5}$
in the SM\ \cite{Lenz:2006hd}.
Since it is often measured in semileptonic decays it is called semileptonic
\textit{CP} asymmetry.

\begin{figure}
\centering
\begin{minipage}{0.64\textwidth}
\includegraphics[width=\textwidth]{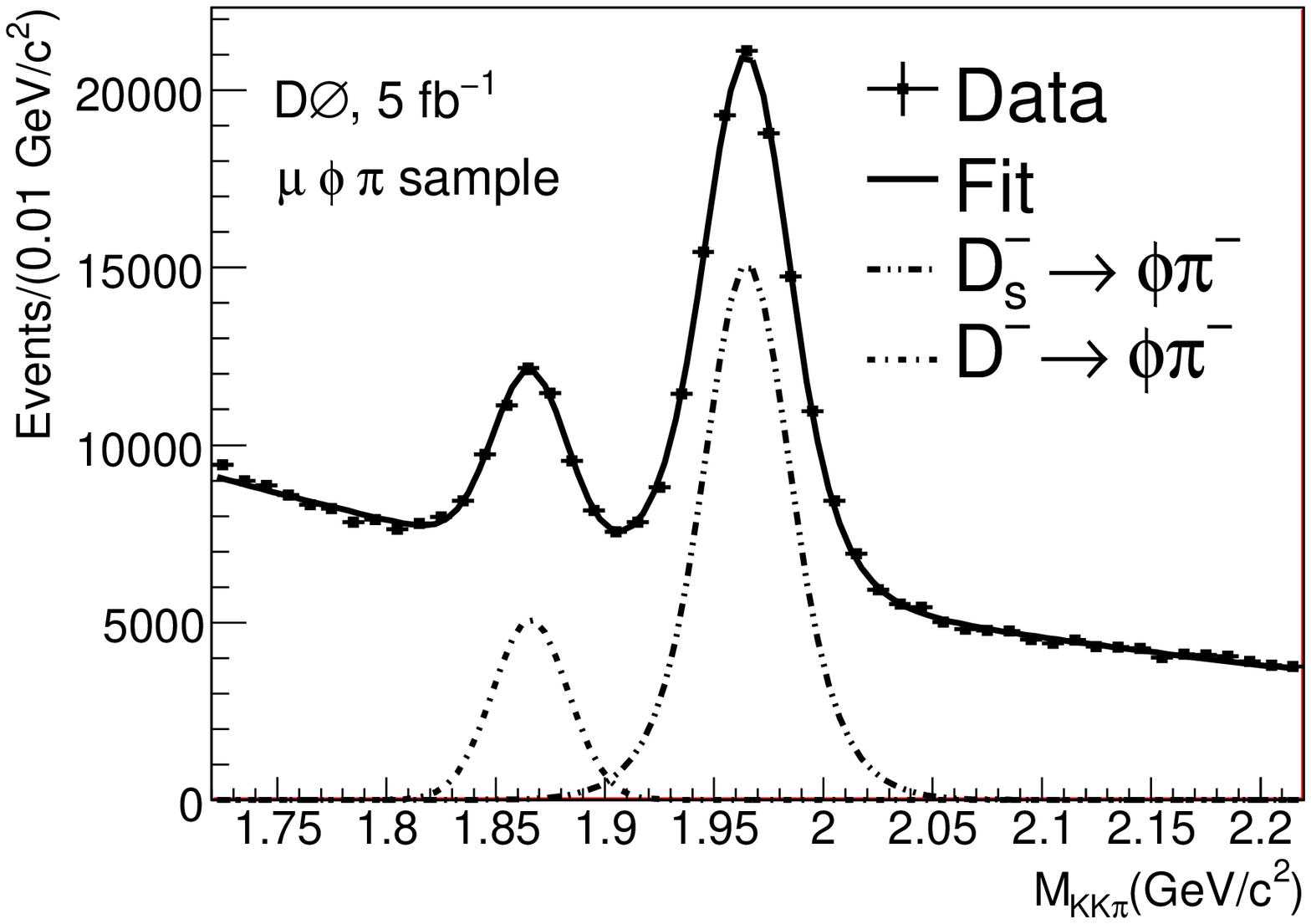}
\end{minipage}
\begin{minipage}{0.33\textwidth}
\includegraphics[width=\textwidth]{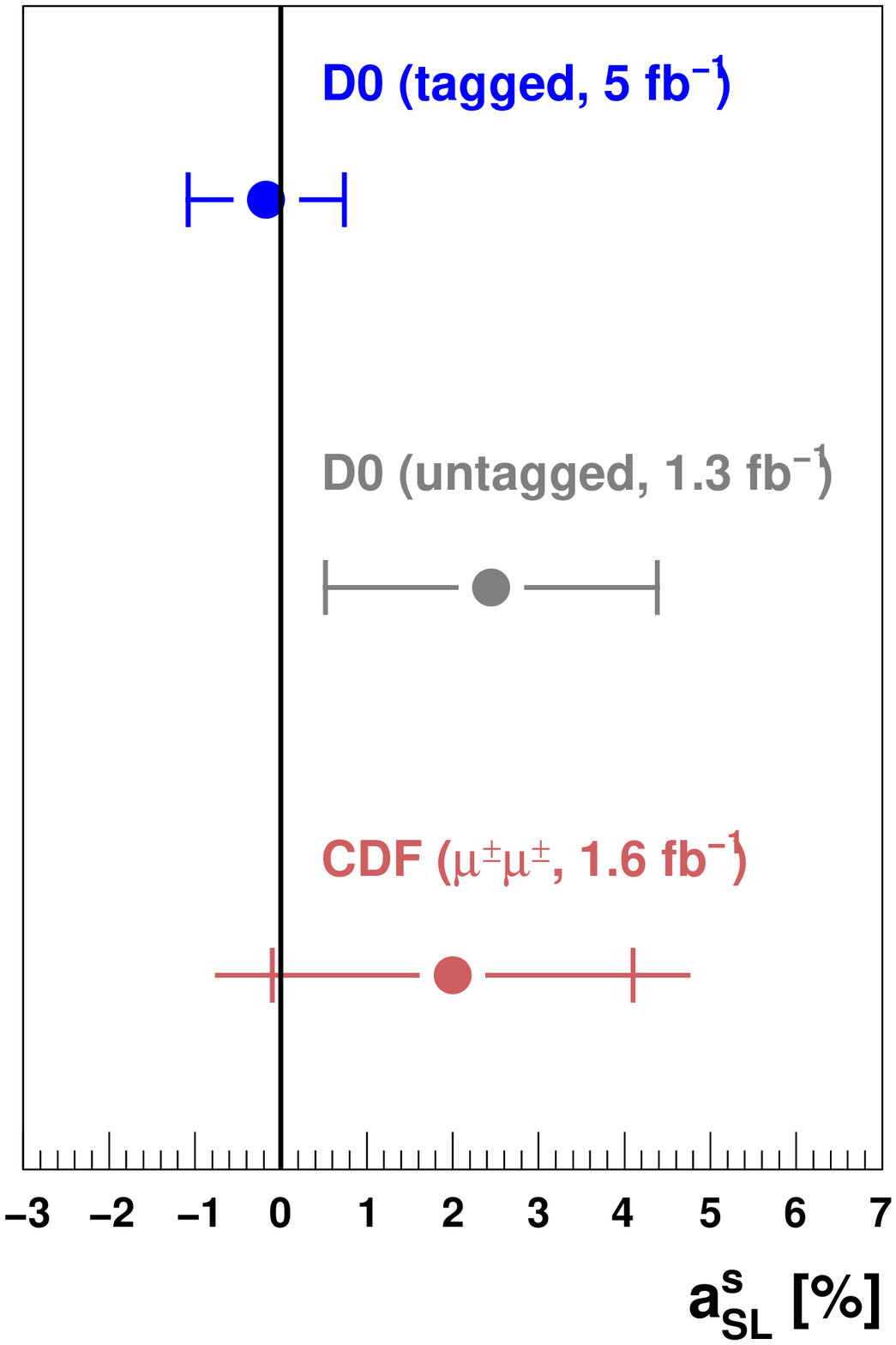}
\end{minipage}
\caption{Invariant mass spectrum of $D^-_s \rightarrow \phi\pi^-$ candidates
measured by D0 (left) and comparison of $a_{SL}$ 
measurements\ \cite{Abazov:2009wg,Abazov:2007nw,CDF:asl} (right).}
\label{fig:aSL}
\end{figure}

D0 measures the asymmetry in $B^0_s \rightarrow \mu^+ D_s^- X$ decays where
the $D_s^-$ is reconstructed in the decays $D_s^- \rightarrow \phi\pi^-$,
$\phi \rightarrow K^+K^-$ (Fig.\ \ref{fig:aSL} left) and $D_s^- \rightarrow K^{*0}K^-$,
$K^{*0} \rightarrow K^+\pi^-$.
The time evolution of $B^0_s$ and $\bar{B}^0_s$ mesons is fitted in an unbinned
maximum likelihood fit where information about the production flavor is determined
from decay products of the second $b$ hadron produced in the collision
(opposite side flavor tagging).
As the final state is not reconstructed exclusively, the missing momentum is
corrected for using simulation.
In a data sample of 5 fb$^{-1}$ D0 measures an asymmetry of 
$a^s_{SL} = (-1.7 \pm 9.1 \mbox{(stat.)} ^{+1.2}_{-2.3} \mbox{(syst.)}) 
\times 10^{-3}$\ \cite{Abazov:2009wg}.
This value agrees well with the SM prediction and previous 
measurements\ \cite{Abazov:2007nw,CDF:asl} as can be seen in
Fig.\ \ref{fig:aSL} right.

\section{\textit{CP} violation in $\mathbf{B^0_s \rightarrow J/\psi\,\phi}$}
\textit{CP} violation in the interference of decays with and without mixing
can be observed in decays to final states that are accessible to $B^0_s$ and $\bar{B}^0_s$.
For the decay to $J/\psi\,\phi$ the phase $\beta_s$ between both processes is in the SM 
given by $\beta_s = \arg(-V_{tb}V_{ts}^*/V_{cb}V_{cs}^*)\approx 0.02$\ \cite{Lenz:2006hd}.
Contributions from new physics could enlarge it in the same way as the phase $\phi$.

As $J/\psi$ and $\phi$ are vector mesons the decay of the pseudoscalar $B^0_s$
can proceed via angular momentum $L=0,1$ or 2, resulting in a \textit{CP} parity of $(-1)^L$.
An angular analysis allows to identify the \textit{CP} eigenstates.
Information about \textit{CP} violation can be extracted from a comparison to the mass eigenstates 
identified by their defined lifetime.
Thus the mixture of final \textit{CP} eigenstates is advantageous if 
$\Delta\Gamma$ is significantly different from zero.

Further sensitivity is achieved by identifying the production flavor (flavor tagging).
It allows to measure the asymmetry
$A_{CP}(t) = [\Gamma(\bar{B}^0_s \rightarrow f)-\Gamma(B^0_s \rightarrow f)]/
[\Gamma(\bar{B}^0_s \rightarrow f)+\Gamma(B^0_s \rightarrow f)]
= \pm\sin(2\beta_s)\sin(\Delta m\, t)$
where the sign depends on the \textit{CP} parity of the final state.
As can be seen from the equation this approach requires to resolve the fast
$B^0_s$ oscillations.

\begin{figure}
\centering
\begin{minipage}{0.42\textwidth}
\includegraphics[width=\textwidth]{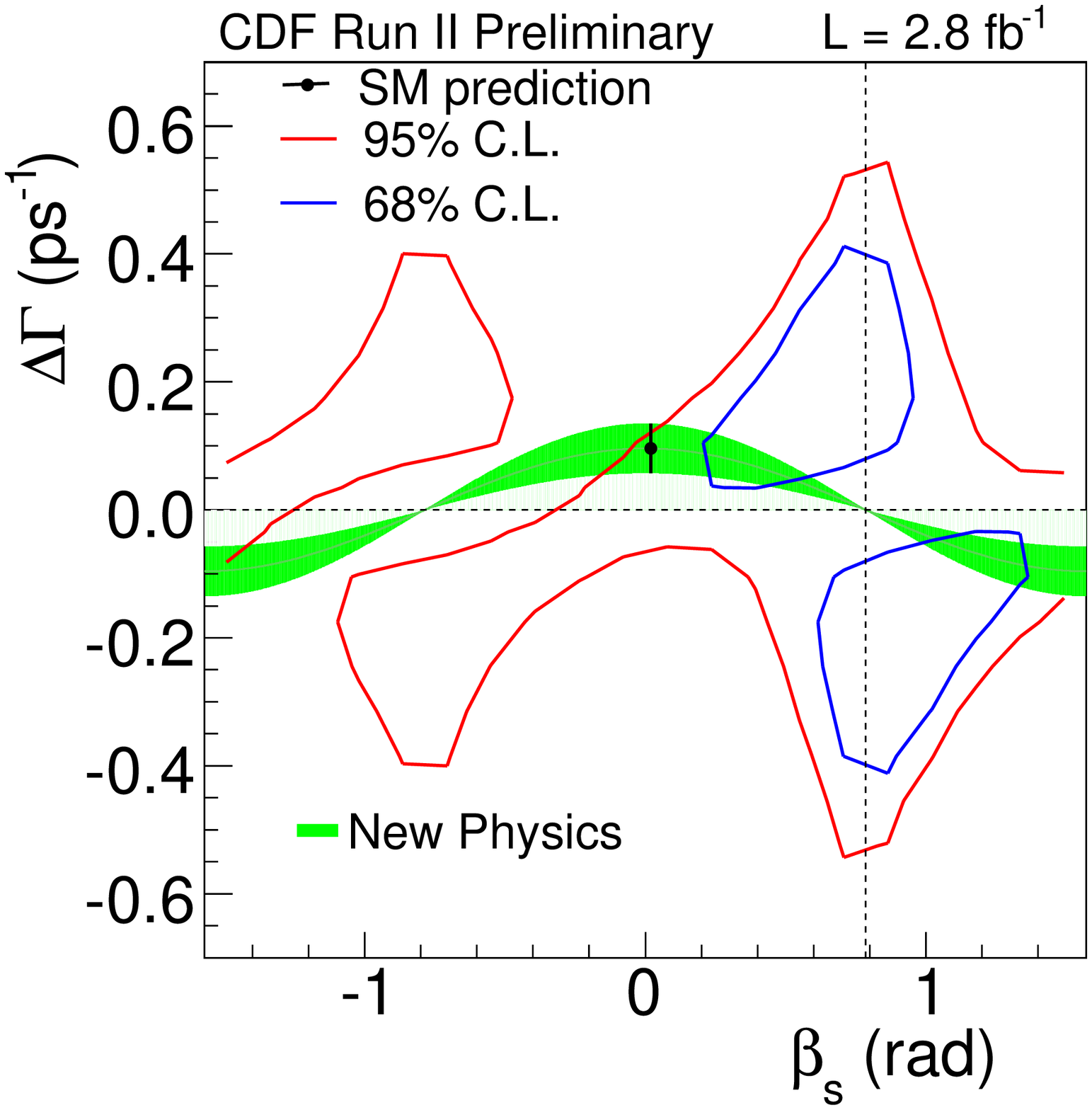}
\end{minipage}
\begin{minipage}{0.53\textwidth}
\includegraphics[width=\textwidth]{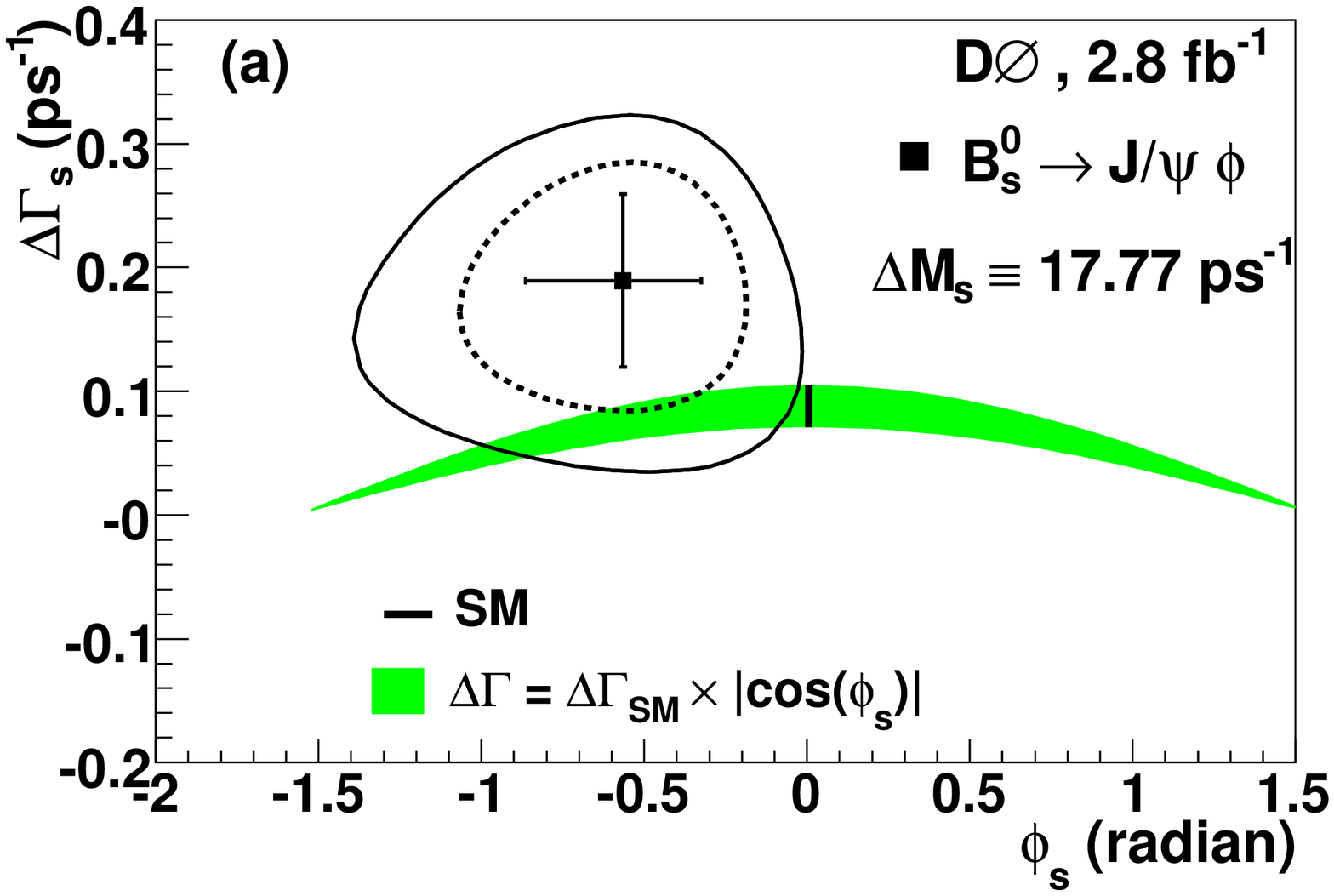}
\end{minipage}
\caption{68\% and 95\% CL regions in the $\beta_s$-$\Delta\Gamma$ plane measured
by CDF (left) and 68\% and 90\% CL regions in the $\phi_s$-$\Delta\Gamma$ plane measured
by D0 (right) where $\phi_s = -2\beta_s$.
The cross shows the $\phi_s$ and $\Delta\Gamma$ values measured by D0.}
\label{fig:Jpsiphi}
\end{figure}

Both Tevatron experiments analyze a data sample of 2.8 fb$^{-1}$ each.
CDF selects about 3200 $B^0_s \rightarrow J/\psi\,\phi$ signal events and 
measures the 68\% and 95\% CL region in the $\beta_s$-$\Delta\Gamma$ plane
shown in Fig.\ \ref{fig:Jpsiphi} left\ \cite{CDF:Jpsiphi}.
Because of unknown strong phases the result exhibits a symmetry with respect to
$\beta_s=\pi/4$ and $\Delta\Gamma=0$.
The one-dimensional confidence region for $\beta_s$ is $[0.28, 1.29]$ at 68\% CL.
D0 selects approximately 2000 signal events and obtains the
confidence regions shown in Fig.\ \ref{fig:Jpsiphi} right\ \cite{Abazov:2008fj}.
From the fit the value 
$\phi_s = -2\beta_s = 0.57 ^{+0.24}_{-0.30}\mbox{(stat.)} ^{+0.07}_{-0.02}\mbox{(syst.)}$ 
is determined.
In this result the strong phases are constrained to values measured in
$B^0 \rightarrow J/\psi\,K^*$ decays, which removes the ambiguity.
In Ref.\ \cite{D0:Jpsiphi} D0 recently presented an updated preliminary result
without constraints on strong phases.

As can be seen in Fig.\ \ref{fig:Jpsiphi} both results are consistent with each other.
The consistency with the SM, measured by the $p$-value, is 7\% (1.8$\sigma$) for CDF
and 8.5\% (1.7$\sigma$) for D0.
The green band in Fig.\ \ref{fig:Jpsiphi} is given by $\Delta\Gamma = 2|\Gamma_{12}|\cos\phi$ 
and shows the region accessible by new physics models that have the same value
of $\Gamma_{12}$, which is dominated by tree level processes, as predicted in the SM.

\section{Conclusions and outlook}
No evidence for physics beyond the SM was found so far in the $B^0_s$ system.
The limits on $\mathcal{B}(B^0_s \rightarrow \mu^+\mu^-)$ place stringent
constraints on new physics models and a further reduction of their parameter space
can be expected from updated measurements.
First evidence or even observation of $B^0_s \rightarrow \mu^+\mu^-\phi$ seems
within reach of the Tevatron.
While the $a^s_{SL}$ measurements agree with the SM within uncertainties,
some deviation from the SM is observed in decays to $J/\psi\,\phi$.
More data is available and will be taken, which gives the Tevatron a realistic
chance to observe new physics if \textit{CP} violation in the $B^0_s$ system is large.


\begin{footnotesize}



%

\end{footnotesize}


\end{document}